\documentclass[preprint2]{aastex}

\newcommand{\swift}{{\it Swift}}

\slugcomment{submitted to PASJ ??, ??.}

\shorttitle{\swift\ publication statistics}
\shortauthors{S.\~Savaglio & U.\~Grothkopf}

\begin{document}


\title{\swift\ publication statistics: a comparison with other major observatories}


\author{S. Savaglio\altaffilmark{1} and U. Grothkopf\altaffilmark{2}}
\affil{}


\altaffiltext{1}{Max Planck Institute for Extraterrestrial Physics, 85748 Garching bei M{\"u}nchen, Germany, savaglio@mpe.mpg.de}
\altaffiltext{2}{European Southern Observatory, 85748 Garching bei M{\"u}nchen, Germany}


\begin{abstract}

{\it Swift} is a satellite equipped with $\gamma$-ray, X-ray, and optical-UV instruments aimed at  discovering, localizing and collecting data from gamma-ray bursts (GRBs). Launched at the end of 2004, this small-size mission finds about a hundred GRBs per year, totaling more than 700 events as of 2012. In addition to GRBs, {\it Swift} observes other energetic events, such as AGNs, novae, and supernovae. Here we look at its success using bibliometric tools; that is the number of papers using \swift\ data and their impact (i.e., number of citations to those papers). We derived these for the publication years 2005 to 2011, and compared them with the same numbers for other major observatories. \swift\ provided data for 1101 papers in the interval 2005-2011, with 24 in the first year, to 287 in the last year. In 2011, \swift\ had more than double the number of publications as Subaru, it overcame Gemini by a large fraction, and reached Keck. It is getting closer to the $\sim400$ publications of the successful high-energy missions XMM-Newton and Chandra, but is still far from the most productive telescopes VLT (over 500) and HST (almost 800). The overall average number of citations per paper, as of November 2012, is 28.3, which is comparable to the others, but lower than Keck (41.8). The science topics covered by {\it Swift} publications have changed from the first year, when over 80\% of the papers were about GRBs, while in 2011 it was less than 30\%. 

\end{abstract}


\section{Introduction}

The scientific impact in its fullest sense of  individual scientists, research institutes, or universities is rather hard to quantify and may take a long time to become apparent. One of the several approaches commonly used, in the absence of a better metric, is bibliometrics, which measures publication and citation numbers. Bibliometrics applies statistical methods to scientific publications, and, as generally done for all subject areas, in astronomy it also measures the impact of a given entity  through the total number of citations to its papers. It is used nowadays by all major astronomical observatories, including HST, VLT, Keck, Chandra, and is generally well documented. HST, ESO telescopes (e.g., VLT), Keck, and VLA have made available their bibliometric statistics beginning in the nineties. The space missions Chandra and XMM-Newton (the high-energy space missions) started their publication statistics more recently, as they were launched later.

All of the mentioned observatories are medium to large-size projects, with budgets above half a billion USD, whose well-proven successful publication statistics are instrumental in advocating support from major space agencies or national and international institutions. Here we present the first complete bibliometric investigation for the years 2005-2011 of the gamma-ray burst (GRB) small-size mission (budget $< 300$ million USD) \swift\ (Gehrels et al.\ 2004). The impact of \swift\ in the astronomical community has already been shown by Madrid \& Macchetto (2009). In their list of the most successful observatories for the year 2006, small projects, including SDSS and \swift, are competing and performing even better than big `monster telescopes' such as HST or VLT.

To perform a meaningful comparison with other observatories in our analysis of publication statistics, \swift\ papers are selected by following the same strict criteria that other major observatories use. First, only peer-reviewed publications are considered. Next, a paper is considered as a \swift\ paper if \swift\ data are used. No theoretical papers (i.e., with no use of data) are included, even if \swift-based results from other publications are mentioned. Papers not considered \swift\ papers include those which only mention an object discovered by \swift, use the coordinates of an object discovered with \swift, or if the work was triggered by a \swift\ discovery. Our work is organized as follows: a short description of the mission is given in \S\ref{mission}. The details of the methodology used are provided in \S\ref{method}. In \S\ref{results}, we describe the results, in \S\ref{science} we discuss the science topics covered by \swift\ publications, and in the final \S\ref{discussion}, we draw the conclusions of our survey.

\section{The Swift mission}\label{mission}

{\it Swift} is a dedicated satellite primarily intended and designed to detect, localize and observe GRBs. Built by NASA and an international consortium from the United States, United Kingdom, and Italy, it was launched in November 2004 and has been operating ever since. It carries three instruments: the Burst Alert Telescope (BAT; Barthelmy et al.\ 2005) is a $\gamma$-ray instrument which detects GRB events and localizes them to arcmin precision; the X-ray Telescope (XRT; Burrows et al.\ 2005) provides a more precise position with an error radius $<2$ arcsec in 90\% of the cases (Evans et al.\ 2009); the 0.3 meter Ultraviolet/Optical Telescope (UVOT; Roming et al.\ 2005) provides sub-arcsec localization when the GRB is optically bright. 

The position is communicated within one minute via the Tracking and Data Relay Satellite System (TDRSS) to the  Mission Operations Center at Penn State University (Pennsylvania) and the Gamma-ray Coordinates Network (GCN) at the Goddard Space Flight Center (GSFC, near Washington, D.C.). The scientific community is publicly alerted for rapid follow up with ground-based telescopes or orbiting satellites, before the blast becomes too faint. Data, which are made public as soon as they are taken, are stored and analyzed at the  \swift\ Science Data Center (at GSFC), at the University of Leicester (UK), and at the ASI Data Center (near Rome). 

\swift\ is operated by $3-4$ scientists working on-line simultaneously, rotating from a team of $14-15$ people. The satellite reached routine observations early on. Considering that BAT has triggers disabled for 18\% of the time due to the South Atlantic Anomaly and during slews, which is an additional 17\% of the time, \swift\ works at a very high efficiency of 98\%. It detects about 100 GRBs every year and, as of end of 2012, has produced data for more than 700 GRBs. The total budget covering launch and the first eight years of operation ($\sim6$ million USD per year) has been about 300 million USD, including contributions from outside the US.  The costs also include about 1.5 million USD per year awarded as grants to US scientists through a Guest Investigator Program. Thanks to its recognized success, NASA has recently granted financial support until 2014.

\begin{table}
\caption[t0]{\swift\ publications and citations in the years 2005-2011\label{t0}}
\begin{center} 
\begin{tabular}{cccc}
\tableline\tableline&&&\\[-5pt] 
Year & \#papers & \#citations & Cites/paper \\
[5pt]\tableline&&&\\[-5pt] 
2005 & 24	 & 2165 & 90.2 \\
2006 & 81	 & 4976 & 61.4 \\
2007	 & 152 & 6204 & 40.8 \\
2008	 & 143 & 4070 & 28.5 \\
2009	 & 190 & 5846 & 30.8 \\
2010 & 224 & 4020 & 17.9 \\
2011 & 287 & 3860 & 13.4 \\
[2pt]\tableline
All & 1101 & 31141 & 28.3 \\
[2pt]\tableline
\end{tabular}
\tablecomments{Columns: (1) Year of publication. (2) Number of \\
\swift\ peer-reviewed publications. (3) and (4) Total and \\
average number of citations, obtained from ADS as of 14 \\
November 2012. }
\end{center}
\end{table}

\begin{figure}
\epsscale{1.02}
\plotone{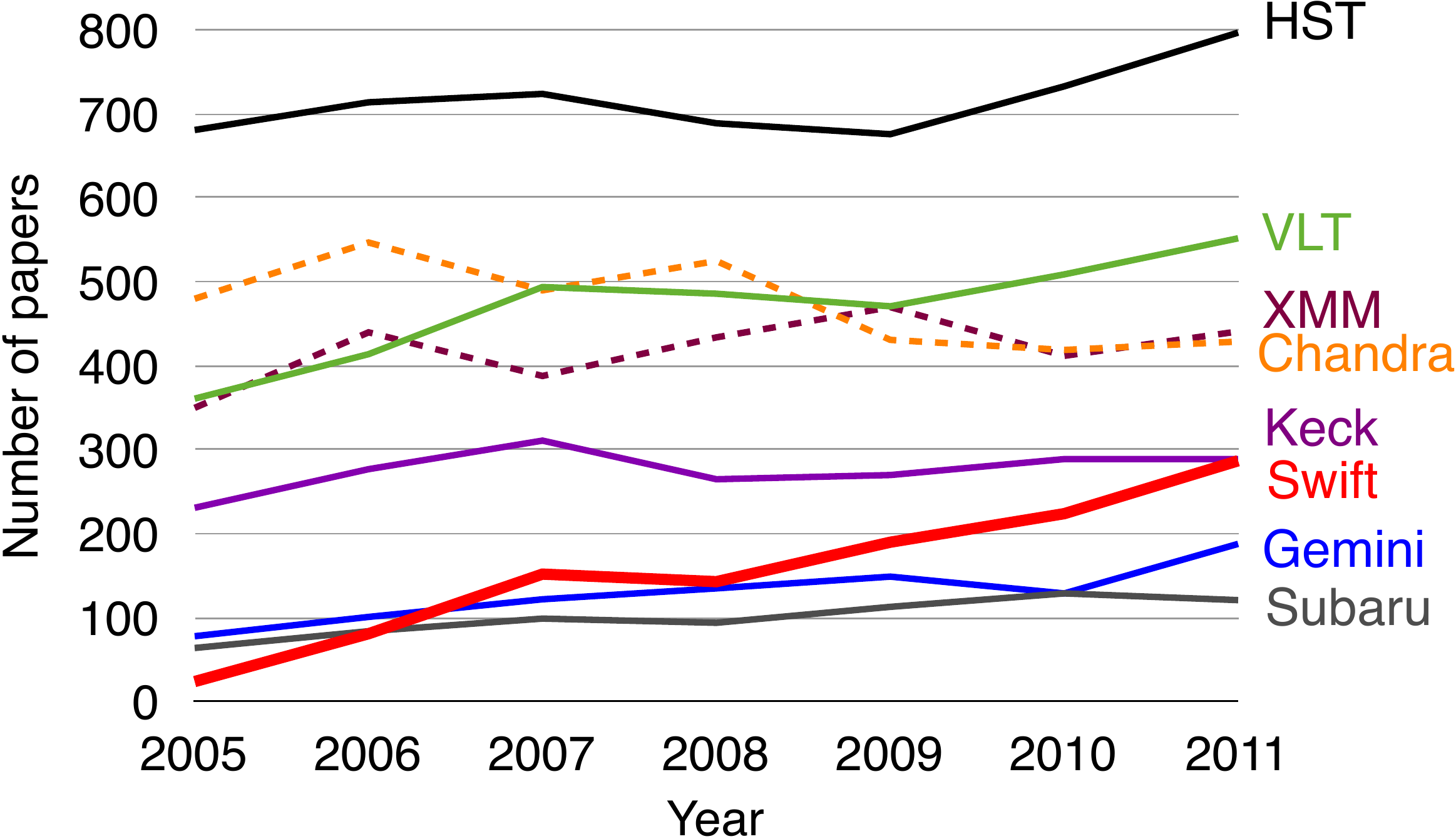}
\caption{Number of papers published with data from different observatories, for the years 2005-2011. {\it Swift} data were first published in 2005 (thick red line), a few months after its launch in November 2004. Chandra and XMM-Newton are represented as dashed lines because publications are counted using different criteria, resulting in general in a higher number of publications.\label{f1}}
\end{figure}

\begin{figure}
\epsscale{1.02}
\plotone{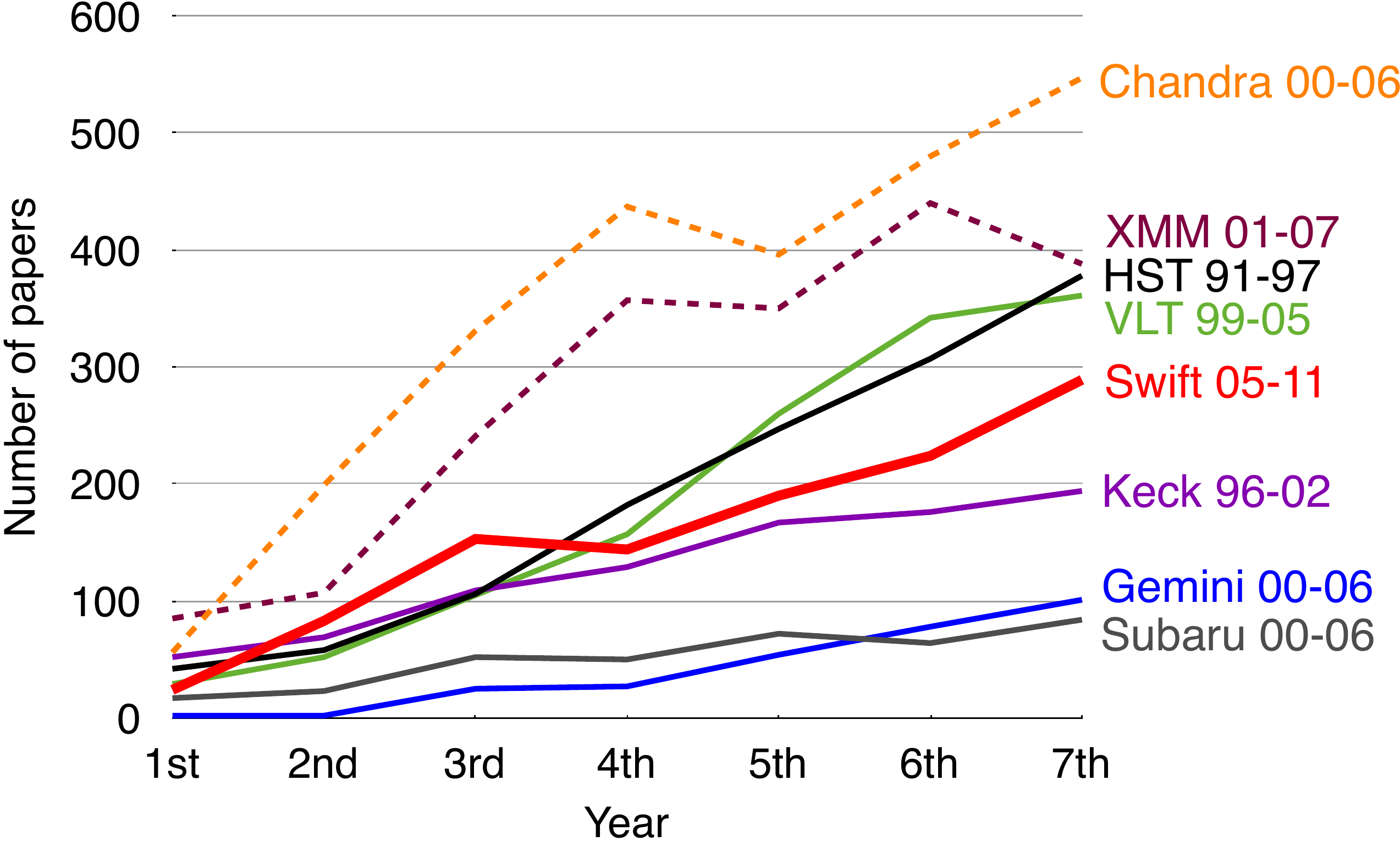}
\caption{Number of papers starting from the first year of publication (indicated in the labels on the right) for each observatory, and for the following 6 years. For \swift\ (thick red line), this corresponds to the interval 2005-2011. Chandra and XMM-Newton are represented as dashed lines because  publications are counted using different criteria, which give generally a higher number of publications.\label{f2}}
\end{figure}

\section{Methodology}\label{method}

The methodology we applied to identify \swift\ papers consists of two parts. First, we use a Full-Text-Search tool (FUSE) to spot papers that mention the word 'Swift' in the text, then we visually inspect each paper to classify it. The peer-reviewed journals we screened are {\it A\&A, A\&ARv, AJ, ApJ, ApJS, AN, ARA\&A, EM\&P, Icarus, MNRAS, Nature, NewA, NewAR, PASJ, PASP, P\&SS} and {\it Science}.

\subsection{FUSE}

FUSE is a program developed and maintained by the European Southern Observatory (ESO) Library. A description can be found in \citet{erd10} and \cite{meak11}. FUSE is  used by telescope bibliography compilers of several major observatories, including HST, Gemini, and Subaru. 

The program relies heavily on the NASA Astrophysics Data System (ADS) Abstract Service. FUSE converts PDFs of specific journals into text files, scans them for user-defined  keywords, and provides highlighted results in context. The sole keyword searched for in our study was {\it Swift}. 

\begin{figure}
\epsscale{1.}
\plotone{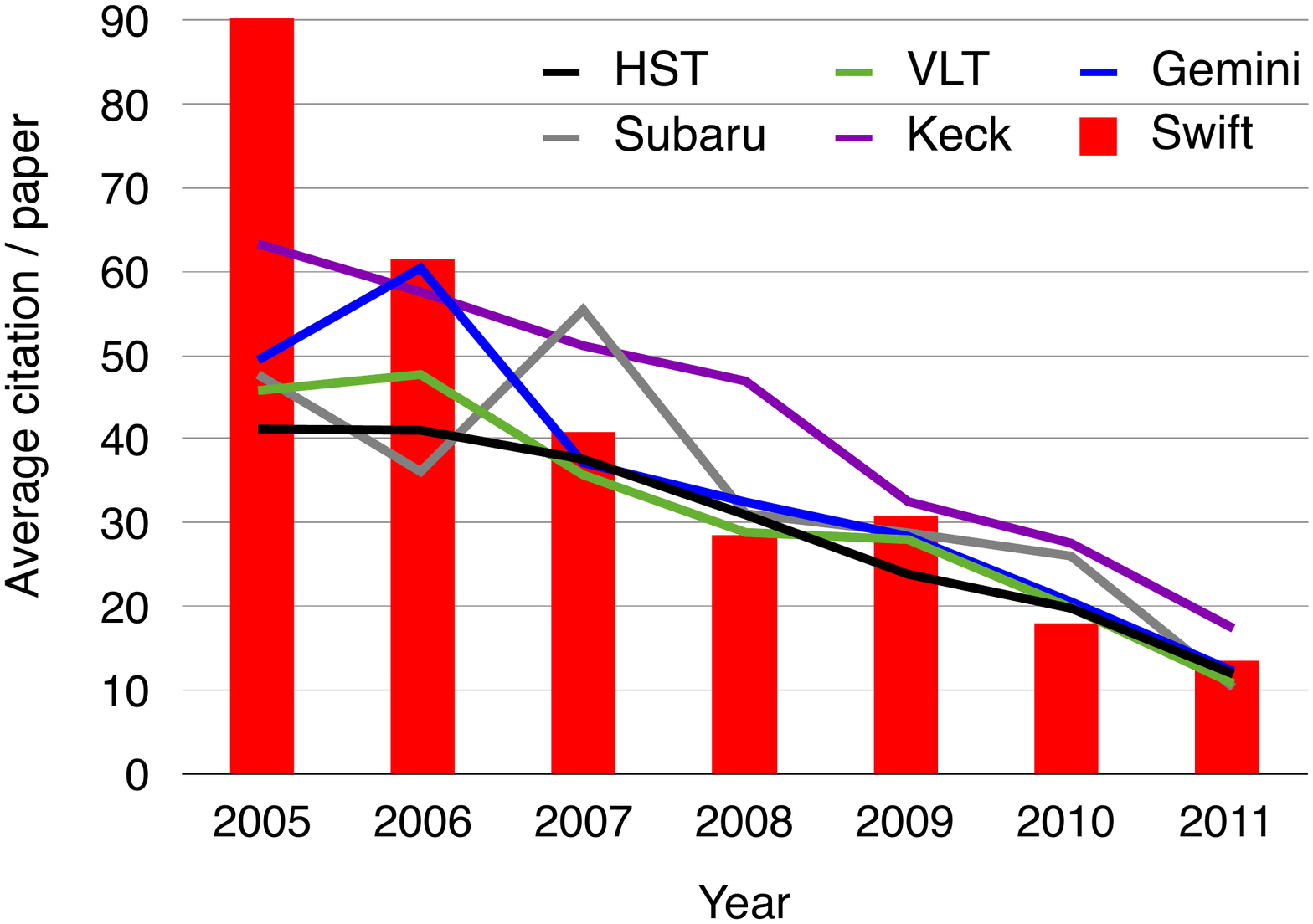}
\caption{Number of citations per paper for \swift\ (red histogram), HST, VLT, Gemini, Subaru and Keck, for papers published 2005-2011 as of mid-November 2012).  The citations build up over time for older papers, so the trend is naturally downwards for more recent papers.\label{fcitations}}
\end{figure}

\subsection{Visual inspection and classification}\label{selection}

Papers selected with FUSE that contain the term \swift\ anywhere in the text were inspected visually to determine whether \swift\ data were actually used for the research or whether the term was mentioned in another context (e.g., future observations with \swift). Papers which use \swift\ coordinates as finding charts or as the basis for observations with other facilities are not included. Likewise, papers which merely quote results presented in other papers are excluded. Finally, papers discussing \swift\ instruments and calibrations (we found 8 in total) are excluded too. About $20-50$\% of the examined papers from 2005 to 2011 are  included in the \swift\ bibliography. When we identify a \swift\ data paper, we also check which instruments were used and keep track of the scientific subject covered. An overview of the number of \swift\ papers, along with the total and average citations (as of 14 November 2012), is given in Table\,\ref{t0}. 

\subsection{Other observatories}\label{observatories}

For ground-based optical/NIR publication statistics, we included those provided for Gemini\footnote{www.gemini.edu/science/publications}, Keck\footnote{www2.keck.hawaii.edu/library/keck\_papers.html}, Subaru\footnote{www.naoj.org/Observing/Proposals/Publish}, and VLT\footnote{www.eso.org/libraries/telbib\_pubstats\_overview.html}. For UV and X-ray satellites, we used HST\footnote{archive.stsci.edu/hst/bibliography}, Chandra\footnote{cxc.harvard.edu/cda/bibstats/bibstats.html} and XMM-Newton\footnote{heasarc.gsfc.nasa.gov/docs/xmm/xmmbib.html.  Number of publications per year have been provided by Norbert Schartel, ESA, Madrid, Spain.} publication statistics. For more details, see Apai et al.\ (2010) on HST bibliometrics, Rots et al.\  (2012) on Chandra, and Grothkopf \& Meakins (2012) on ESO telescopes (including VLT). For the ground-based telescopes and HST, statistics are obtained using the same rigorous method described in \S\ref{selection}, therefore results can be considered comparable with \swift's. For Chandra and XMM-Newton, a different methodology is used, giving generally higher numbers. For example, a paper could describe models or a theory which are based on Chandra results published in another paper. According to Chandra publication statistics, both papers are  Chandra papers, whereas, according to our method, only the original paper would qualify. For \swift, HST and all ground-based telescopes, publication statistics are limited to data use only.

\begin{table*}
\caption[t0]{Number of \swift\ publications in the years 2005-2011 per journal\label{tjournals1}}
\begin{center} 
\begin{tabular}{cccccccccc}
\tableline\tableline&&&&&&&&&\\[-5pt] 
Journal & 2005 & 2006 & 2007 & 2008 & 2009 & 2010 & 2011 & Total & \% \\
[5pt]\tableline&&&&&&&&&\\[-5pt] 
ApJ/ApJS	& 14	& 46	& 78 & 74 & 90	& 103 & 127	& 532 & 48.3	\\
A\&A 	& 2	& 19	& 49 & 37	& 37	& 41	& 67		& 252 & 22.9	\\
MNRAS	& 1	& 9	& 18 & 20	& 53	& 70	& 75		& 246 & 22.3	\\
AJ		& 0 	& 1	& 4	& 4	& 6	& 4	& 3		& 22	 & 2.0	\\
PASP	& 0 	& 0	& 1	& 0	& 0	& 1	& 0		& 2	& 0.2		\\
Nature	& 6 	& 6	& 0	& 5	& 3	& 0	& 5		& 25 & 2.3		\\
Science	& 1 	& 0	& 0	& 1	& 0	& 2	& 4		& 8	& 0.7		\\
Others	& 0	& 0	& 2	& 2	& 1	& 3	& 6		& 14 & 1.3		\\
[2pt]\tableline
\end{tabular}
\end{center}
\end{table*}

\begin{table}
\scriptsize
\caption[t0]{\swift, VLT and HST publications (2005-2011) per journal\label{tjournals2}}
\begin{center}
\begin{tabular}{ccccccccc}
\tableline\tableline&&&&&&&&\\[-5pt]
 &  \multicolumn{2}{c}{\swift} && \multicolumn{2}{c}{VLT\tablenotemark{a}} && \multicolumn{2}{c}{HST\tablenotemark{b}} \\
[4pt]  \cline{2-3} \cline{5-6} \cline{8-9} \\[-4pt] 
Journal & Tot & \% && Tot & \% && Tot & \% \\
[5pt]\tableline&&&&&&&&\\[-5pt] 
ApJ/ApJS	&  532 & 48.3	&& 824	& 25.1	&& 2407	& 48.3 \\
A\&A 	&  252 & 22.9	&& 1623	& 49.4	&& 755	& 15.1 \\
MNRAS	& 246 & 22.3	&& 591	& 18.0	&& 775	& 15.5 \\
AJ		&  22	 & 2.0	&& 91	& 2.8		&& 493	& 9.9 \\
PASP	&  2	& 0.2		&& 16	& 0.5	 	&& 72	& 1.4	 \\
Nature	& 25 & 2.3		&& 42	& 1.3		&& 52	& 1.0 \\
Science	& 8	& 0.7		&& 14	& 0.4		&& 24	& 0.5  \\
Others	& 14 & 1.3 	&& 86 	& 2.6		&& 406 	& 8.1 \\
[2pt]\tableline
All		& 1101 & --	&& 3287 & --  && 4984 & -- \\
[2pt]\tableline
\end{tabular}
\tablecomments{Columns: (2), (4) and (6) Total numbers of publications \\ for a given journal. (3), (5) and (7) Fraction over the total. }
\tablenotetext{a}{Obtained from ESO Telescope Bibliography (http://telbib.eso.org)}
\tablenotetext{b}{Obtained from HST Bibliography \\ (http://archive.stsci.edu/hst/bibliography/).}
\end{center}
\end{table}

\section{Results}\label{results}


In Fig.\,\ref{f1}, we compare \swift\ papers to publications of major observatories for the years 2005-2011. First \swift\ publications appeared in 2005. All other observatories in this study started earlier, e.g., HST in 1992 and VLT in 1999. Fig.\,\ref{f2} shows publication statistics for early years for all considered observatories. One way to measure the impact of \swift\ is by counting the number of papers which use \swift\ data, and the related number of citations. We did this for the years 2005 to 2011, and compare these numbers with all major observatories listed in \S\ref{observatories}.

\subsection{Number of publications}

\swift\ started successful operations right away, after its launch in September 2004. The first GRB was detected December 17 2004, called GRB~041217. The first \swift\ papers appeared in 2005, the year following its launch, due to the normal delay for the publication process. The number of publications shows an almost steady increase (Fig.\,\ref{f1}), starting with 24 papers in 2005, and increasing to 287 in 2011. In the last year, \swift\ publications were more than those for Gemini, twice those for Subaru, and equalled those of  one of the most successful observatories ever, Keck. The total number of \swift\ publications in 2005-2011 is 1101. Details are reported in Table\,\ref{t0}, whereas Fig.\,\ref{f1} shows the comparison with major observatories. The distribution of journals where these papers were published is shown in Table\,\ref{tjournals1}, and the comparison with VLT and HST in Table\,\ref{tjournals2}. Almost half of the \swift\ papers appeared in the American journal {\it ApJ}, while 45\% were equally distributed in the two European journals {\it A\&A} and {\it MNRAS}. This reflects the multi-national aspect of the mission. The fraction of VLT publications in European ({\it A\&A} and {\it MNRAS}) and American ({\it ApJ, AJ, PASP}) core astronomy journals is 67\% and 28\%, respectively (Table\,\ref{tjournals2}). The situation for HST is reversed, European and American publications are 31\% and 60\%, respectively. The journals with particularly high impact factors, {\it Science} and {\it Nature}, together published 3\% of all \swift\ papers, more than what is typically found for other more mature observatories (1.7\% and 1.5\% for VLT and HST, respectively).

Publication statistics are generally affected, e.g., by the number of telescopes (VLT has four) and instruments of an observatory, hours of observation, construction costs and maintenance, and the number of years of operation (HST started observations more than 20 years ago). Moreover, ground-based telescopes are affected by poor weather conditions, variable seeing and bright time, as well as by the fact that they cannot observe during the day. Publications based on data from any observatory typically start the year following the {\it first light}. In Fig.\,\ref{f2} we display the number of publications for the first 7 years of publication for the observatories considered here. The steady increase is common to all. This is related to the organization of observatories, with instruments coming online successively during the first years. A steady regime for \swift\ has likely not been reached yet as of 2011.
For 2012, preliminary predictions indicate over 350 papers, getting closer to the larger high-energy missions XMM-Newton and Chandra, which in 2011 published about 450 papers.

\begin{figure*}
\epsscale{1.65}
\plotone{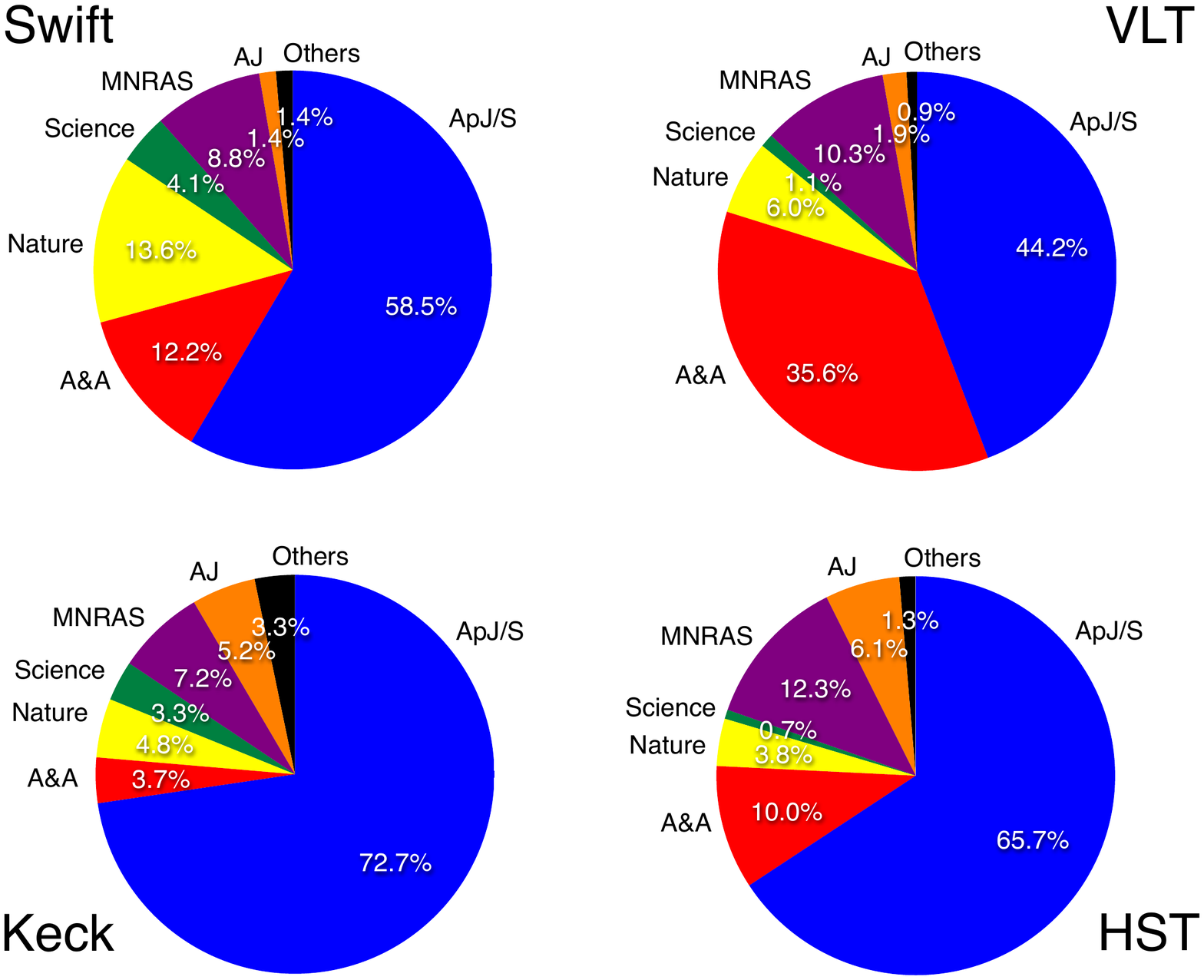}
\caption{Distribution per journal and for different observatories of the most successful publications, in 2005-2011. Most successful means the top 5\% in terms of citations of all publications. Citations obtained in November 2012 from ADS and the ESO Library. \label{fivepercent}}
\end{figure*}

\subsection{Impact: citations} \label{citations}

A key parameter in evaluating the success of an observatory is the number of citations to its papers for each year or for all years. Of course the number of citations to papers of a given year is a dynamical measure, as it keeps growing with time. In Fig.\,\ref{fcitations}, the number of citations (as measured in mid-November 2012, provided by the ESO Library for VLT and obtained from ADS for all other telescopes) relative to publications of 2005-2011 is reported for \swift\ and other major observatories. The highest number of citations are obviously attached to the oldest papers, because they have had more time to accumulate citations. The first year of publication was particularly successful, with a very high number of average citations per paper of $\sim90$. Among the 24 papers published in 2005, 6 were in {\it Nature} and one in {\it Science}, which helps explain the high citation number. The overall average number of citations per paper (the total number of citations divided by the total number of papers) is 28.3, which is comparable to the others (HST: 29.2, VLT: 29.6, Gemini: 31.2, Subaru: 31.8), but lower than Keck (41.8). Among Keck publications, several are very successful (which results in a higher average number of citations), while the citations of other observatories are more uniformly distributed.

\begin{table*}
\scriptsize
\caption[t1]{Number of high impact \swift\ papers for the years 2005-2009\label{t1}}
\begin{center} 
\begin{tabular}{ccccccccccccccccc}
\tableline\tableline&&&&&&&&&&&&&&&&\\[-5pt] 
 & &  \multicolumn{3}{c}{\swift} && \multicolumn{3}{c}{VLT} && \multicolumn{3}{c}{HST} &&\multicolumn{3}{c}{Keck} \\
[4pt]  \cline{3-5} \cline{7-9} \cline{11-13}\cline{15-17} \\[-4pt] 
Year & Total papers & \# & HIPs & \% &&  \# & HIPs & \% && \# & HIPs & \% &&  \# & HIP & \% \\
 & ($\times 10^3$) & papers &  &  &  &  papers &  &   && papers & & && papers  &  &  \\
[5pt]\tableline&&&&&&&&&&&&&&&&\\[-5pt] 
2005 & 24.8	 	& 24	& 4	& 16.7	&& 361	& 7	& 1.9 	&& 681	& 13	& 1.9 	&& 231	& 13	& 5.9 \\
2006 & 25.3	 	& 81	& 6	& 7.4		&& 414	& 16	& 3.9 	&& 714	& 25	& 3.5 	&& 277	& 14	& 5.2 \\
2007	 & 23.3	 	& 152 & 3	& 2.0		&& 494	& 11	& 2.2 	&& 724	& 27	& 3.7 	&& 311	& 16	& 5.1 \\
2008	 & 26.2	 	& 143 & 3	& 2.1		&& 486	& 10	& 2.1 	&& 689	& 17	& 2.5 	&& 265	& 14	& 5.3 \\
2009	 & 27.3		& 190 & 5 & 2.6		&& 471	& 15	& 3.2		&& 676	& 15	& 2.2 	&& 270	& 12	& 4.4 \\
[2pt]\tableline
\end{tabular}
\tablecomments{Columns: (2) Total number of peer-reviewed publications, according to ADS, in units of 1000. (3),(6), (9) and (12) Number of papers for the given observatory. (4), (7), (10) and (13) Number of high impact papers. (5), (8), (11) and (14) Fraction of high impact papers over the total.}
\end{center}
\end{table*}

Another way of measuring the success of publications is by considering the number of so-called {\it high impact} papers (HIPs).\footnote{http://thomsonreuters.com/products\_services/science/\\science\_products/a-z/high\_impact\_papers/} 
According to the company Thomson Reuters (producers of the Science Citation Index, SCI), HIPs of a given year are the top 200 most cited papers for that year. These can also be found by using ADS, and include only refereed papers, generally between $\sim24,800$ to $\sim27,300$ in total in the years 2005-2009 (Table\,\ref{t1}).  We counted the number of papers among the top 200 most cited ones  for \swift, HST, VLT and Keck for these years (we do not show the results for the years 2010 and 2011, due to a lack of significant statistics). Results are in Table\,\ref{t1}. In the first year of publication for \swift, 4 out of 24 papers (almost 17\%) are HIPs, which is much more than the total fraction for the following 4 years (3.0\%). For VLT and HST, the total fraction is 2.7\% and 2.8\%, respectively. Keck, as already mentioned in the previous paragraph, is a very successful observatory, with an average fraction of HIPs of 5.1\%.

Finally, we extended the concept of successful publications by considering the top 5\% based on the number of citations (as given by  ADS in November 2012). For the years 2005-2011, the top 5\% in terms of citations correspond to approximately the top 1,200-1,300 papers. Fig.\,\ref{fivepercent} shows how these are distributed among journals, in a cumulative way for the years 2005-2011. The fraction of papers in the high-impact journals {\it Nature} and {\it Science} is 18\% for \swift, while the same fraction for the other observatories are in the range 5\%-8\%. This is perhaps expected, because \swift\ is newer than the other observatories we considered. We notice that the distribution of the most successful papers for \swift, VLT and HST for different journals shown in Fig.\,\ref{fivepercent} differs from the overall distribution for these observatories for each journal reported in Table\,\ref{tjournals2}. For instance, the fraction of {\it Science} and {\it Nature} papers increases from 3\% in all \swift\ publications (Table\,\ref{tjournals2}) to 17.7\% for the most successful publications. In contrast, {\it A\&A} papers for all \swift, VLT and HST publications decreases from 15\%-49\% to 10\%-36\% when only the most successful publications are considered.

\begin{table*}
\scriptsize
\begin{center}
\caption{The most cited \swift\ papers for the years 2005-2009\label{t2}}
\begin{tabular}{clllcc}
\tableline\tableline
\# & Authors & Bibcode & Title & \#Citations\tablenotemark{a} & Instruments \\
\tableline
& &  2005 &  & & \\
\tableline
1 &  Gehrels et al. & 2005Natur.437..851G & A short $\gamma$-ray burst apparently associated & 303 & BAT/XRT/UVOT \\
   &                            &  & with an elliptical galaxy at redshift $z = 0.225$ & & \\

2 & Burrows et al. & 2005Sci...309.1833B & Bright X-ray Flares in Gamma-Ray Burst & 291 & BAT/XRT/UVOT \\ 
   &  &  & Afterglows &  & \\
3 & Fox et al. & 2005Natur.437..845F & The afterglow of GRB 050709 and the   & 278 & XRT \\
  &  & & nature of the short-hard $\gamma$-ray bursts & & \\
4 & Barthelmy et al. & 2005Natur.438..994B & An origin for short $\gamma$-ray bursts & 212 & BAT/XRT \\
   &                &  & unassociated with current star formation &  & \\
5 & Tagliaferri et al. & 2005Natur.436..985T & An unexpectedly rapid decline in the X-ray & 172 & BAT/XRT \\
   &                   &    &  afterglow emission of long $\gamma$-ray bursts & & \\
\tableline
 & & 2006  & & & \\
\tableline
1 &  Nousek et al. & 2006ApJ...642..389N & Evidence for a Canonical Gamma-Ray Burst  & 455 & XRT  \\
   &                            &  & Afterglow Light Curve in the Swift XRT Data & & \\
2 & Campana et al.  & 2006Natur.442.1008C  & The association of GRB 060218 with a  & 342 & BAT/XRT/UVOT \\
   &  &  &  supernova and the evolution of the shock wave   &  & \\
3 & O'Brien et al. & 2006ApJ...647.1213O & The Early X-Ray Emission from GRBs & 254  & BAT/XRT/UVOT \\
4  & Gehrels et al. & 2006Natur.444.1044G  & A new $\gamma$-ray burst classification scheme from  & 213 & BAT \\
     &  &  & GRB060614  &  & \\
5   & Bloom et al. & 2006ApJ...638..354B & Closing in on a Short-Hard Burst Progenitor:   &  206 & XRT \\
     & & & Constraints from Early-Time Optical Imaging & & \\
     & & & and Spectroscopy of a Possible Host  Galaxy & & \\
     & & & of GRB 050509b & & \\
5   & Soderberg et al. & 2006Natur.442.1014S & Relativistic ejecta from X-ray flash & 206 & BAT/XRT \\
	& & & XRF 060218 and the rate of cosmic explosions & & \\
\tableline
 & & 2007 & & & \\
\tableline
1   & Schaefer & 2007ApJ...660...16S & The Hubble Diagram to Redshift $>6$ from & 208 & BAT \\
     & & &  69 Gamma-Ray Bursts  & & \\
2 & Evans et al. & 2007A\&A...469..379E & An online repository of Swift/XRT light & 184 & XRT \\
     & & &  curves of $\gamma$-ray bursts & & \\
3 & Willingale et al. & 2007ApJ...662.1093W & Testing the Standard Fireball Model of & 162 & BAT/XRT \\ 
     & & &  Gamma-Ray Bursts Using Late X-Ray  & & \\
     & & & Afterglows Measured by Swift && \\
4 & Albert et al. & 2007ApJ...665L..51A & Very High Energy Gamma-Ray Radiation&144 & BAT \\ 
     & & &  from the Stellar Mass Black Hole Binary & & \\
     & & &  Cygnus X-1 & & \\
5 & Butler et al. & 2007ApJ...671..656B & A Complete Catalog of Swift Gamma-Ray & 142 & BAT \\ 
     & & & Burst Spectra and Durations: Demise of a & & \\
     & & & Physical Origin for Pre-Swift High-Energy && \\
     &&&   Correlations & & \\                    
\tableline
 & & 2008 & & & \\
\tableline
1 & Racusin et al. & 2008Natur.455..183R & Broadband observations of the naked-eye  & 221 & BAT/XRT/UVOT \\ 
     & & & $\gamma$-ray burst GRB080319B & & \\
2 &Tueller et al. & 2008ApJ...681..113T & Swift BAT Survey of AGNs & 163 & BAT/XRT \\ 
3 & Soderberg et al. & 2008Natur.453..469S & An extremely luminous X-ray outburst at the  & 159  & BAT/XRT/UVOT \\ 
     & & & birth of a supernova & & \\
4 & Sakamoto & 2008ApJS..175..179S & The First Swift BAT Gamma-Ray Burst  & 103 & BAT/XRT \\
   & & & Catalog & & \\
5  & Liang et al. & 2008ApJ...675..528L & A Comprehensive Analysis of Swift XRT  & 85 & XRT \\
    & & & Data.\,III Jet Break Candidates in X-Ray and   &  & \\
    & & & Optical Afterglow Light Curves & & \\
5  &  Prieto et al. & 2008ApJ...681L...9P & Discovery of the Dust-Enshrouded Progenitor  & 85 & UVOT \\ 
     & & & of SN 2008S with Spitzer & & \\
\tableline
 & & 2009 & & & \\
\tableline
1 & Tanvir et al. & 2009Natur.461.1254T & A $\gamma$-ray burst at a redshift of $z\sim8.2$ & 240  & BAT/XRT \\ 
2 & Evans et al. & 2009MNRAS.397.1177E & Methods and results of an automatic analysis & 224 & BAT/XRT/UVOT \\ 
     & & &  of a complete sample of Swift-XRT  & & \\
     & & & observations of GRBs & & \\
3 & Salvaterra et al. & 2009Natur.461.1258S & GRB090423 at a redshift of $z\sim8.1$ & 217 & BAT/XRT \\ 
4 & Fynbo et al. & 2009ApJS..185..526F & Low-resolution Spectroscopy of Gamma-ray  & 126 & BAT/XRT \\ 
     & & & Burst Optical Afterglows: Biases in the Swift & & \\
     & & & Sample and Characterization of the Absorbers & & \\
5 & Greiner et al. & 2009ApJ...693.1610G & GRB 080913 at Redshift 6.7 & 123 & XRT \\ 
\tableline
\end{tabular}
\tablenotetext{a}{Number of citations as of November 2012.}
\end{center}
\end{table*}

\section{Science topics}\label{science}

\begin{figure}
\epsscale{1}
\plotone{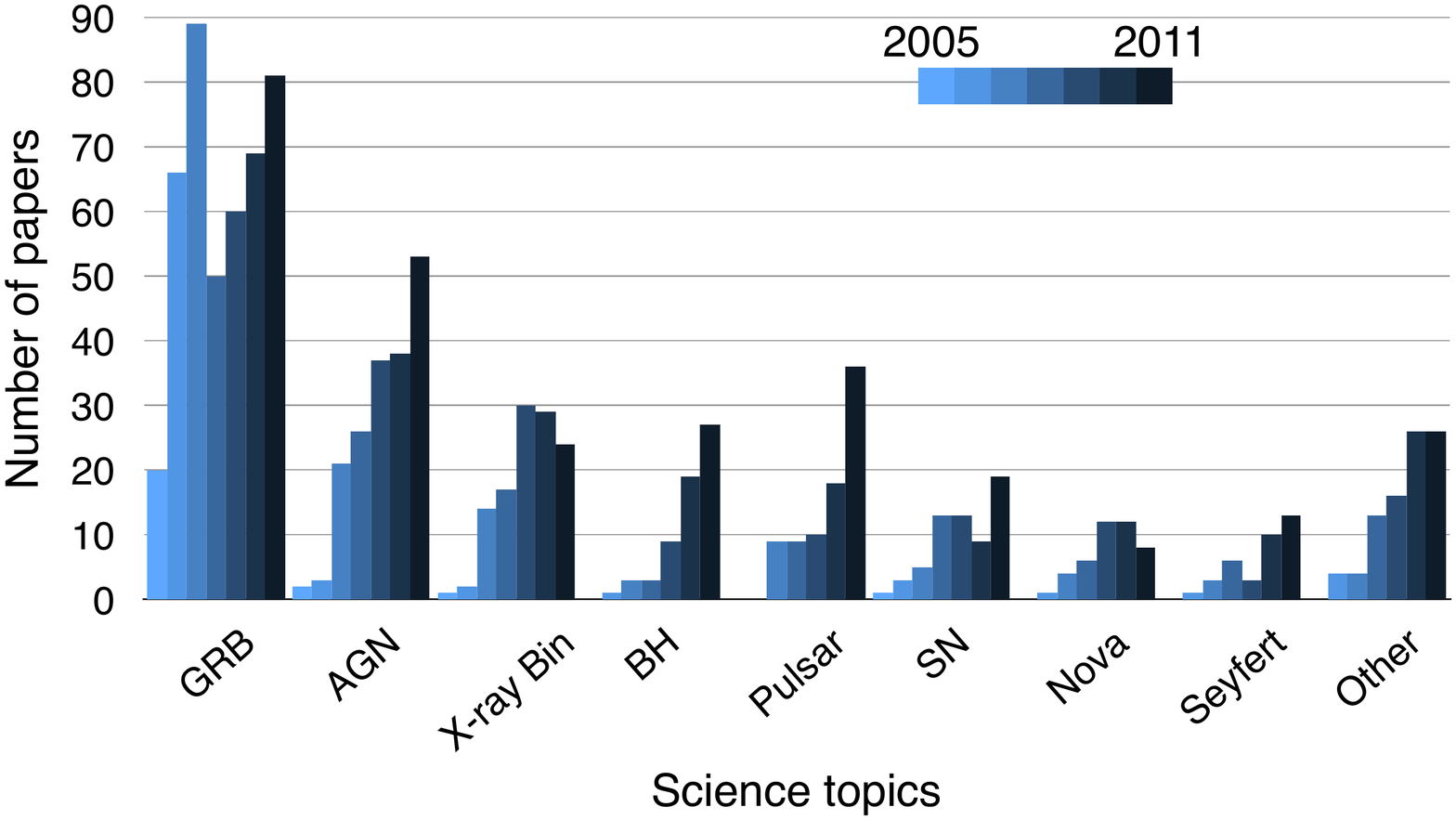}
\caption{Number of papers for \swift\ ordered according to the science topic, for the years 2005-2011. An AGN paper is also dealing with distant QSOs, BL Lacs or blazers. BH indicates massive or stellar black holes. SN includes supernova and supernova remnant. \label{topics}}
\end{figure}

\begin{figure}
\epsscale{1}
\plotone{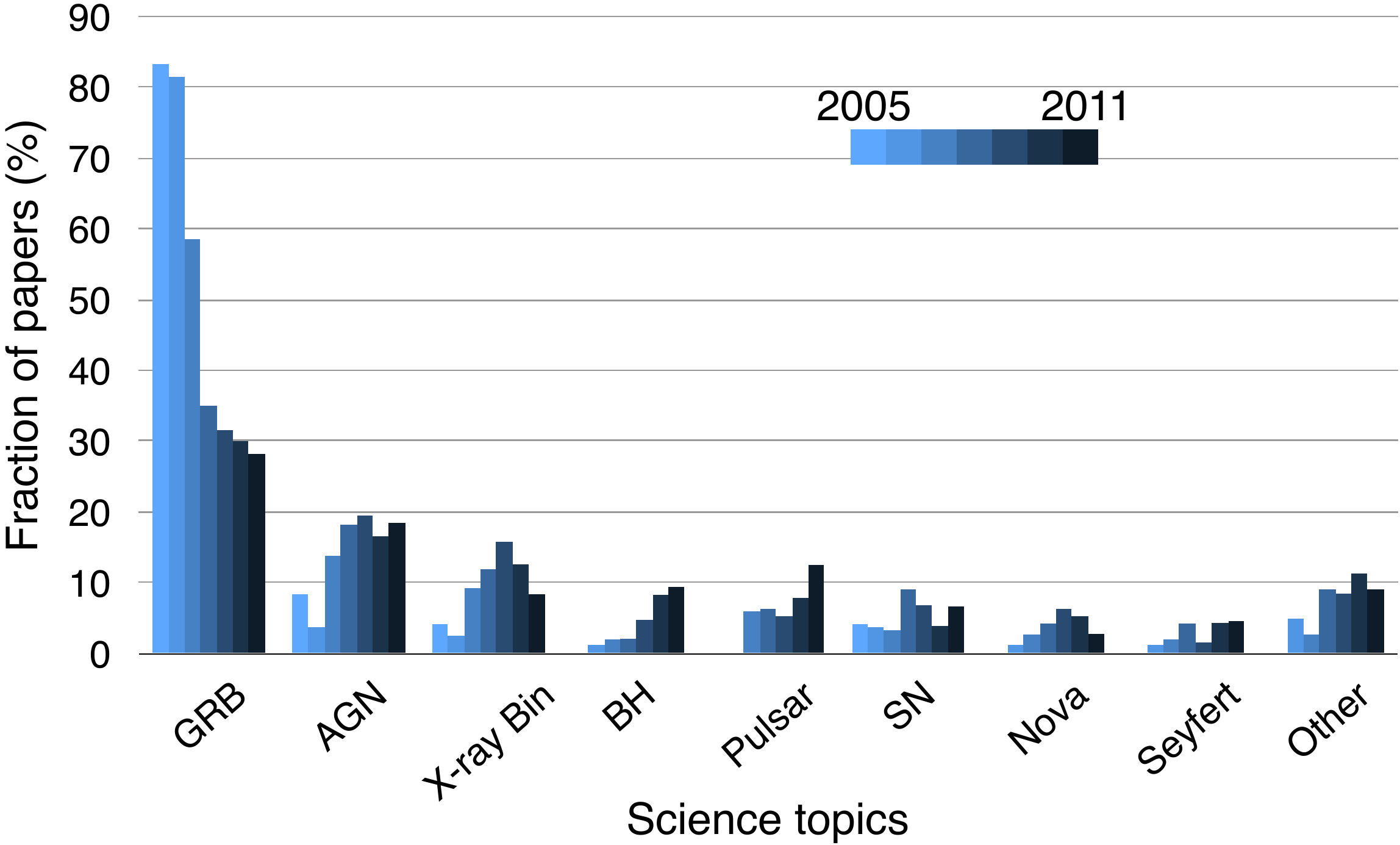}
\caption{Fraction of papers over the total for \swift\ ordered according to the science topic, for the years 2005-2011. Labels are as in Fig.\,\ref{topics}.\label{ntopics}}
\end{figure}

While the main aim of \swift\ is to detect GRBs, the satellite also successfully observes other energetic events.  This is demonstrated by the distribution of the absolute number (Fig.\,\ref{topics}) and the fraction (Fig.\,\ref{ntopics}) of papers for the different science topics. It is clear that in the first two years GRB science was dominant, accounting for over 80\% of the publications. In the following year, 2007, the majority are still GRB papers, almost 60\%. During the years 2008-2011, the science done by \swift\ is mostly non-GRB (which drops to a range of 28\%-35\%). A large fraction are dedicated to galactic sources: $\sim 1/4$ of the papers in 2008-2011 are about X-ray binaries, pulsars, supernovae. Among all topics, after GRBs, the most popular is `AGN', which in our classification includes AGNs, distant QSOs, BL Lac and blazars. \swift\ produced interesting results on many other subjects which are summarized in Figs.\,\ref{topics} and \ref{ntopics} as `Other'. These are mainly extragalactic papers, including X-ray background radiation, galaxy surveys,  galaxy clusters and groups, and radio galaxies. Five papers (out of 1101) are dealing with X-ray and UV emission of comets. It is notable that the absolute number of publications about GRBs in the interval considered, 2005-2011, is not obviously changing,  i.e., the GRB publication rate is not declining in absolute terms. 

Table\,\ref{t2} lists the details of the most successful \swift\ papers per year, for 2005-2009, as indicated by the citation count. The science topic covered by the majority, 21 out of 27 papers, is GRB. One is about AGN, two about supernovae, and one about the black hole in the X-ray galactic source Cygnus X-1. Out of the 27 papers, 11 are published in {\it Nature}, one in {\it Science}. Three (11.1\%) are papers with a female first author (Soderberg et al. 2006; Racusin et al. 2007; Soderberg et al. 2008). Although this number is not particularly significant (large statistical error), it is relatively high.  For comparison, in a recent investigation, Conley \& Stadmark (2012) pointed out that women commissioned to write articles in {\it News \& Views} in {\it Nature} and {\it Perspectives} in {\it Science} for 2010 and 2011 were under-represented. For instance, the proportion for physical sciences articles is 8.1\%, whereas, the proportion of female scientists in the same disciplines in the United States is much higher, 16\% (but much lower, when considering full professors or high-level positions).

In Table\,\ref{tinstruments} we list the number of \swift\ papers distributed according to the use of the three instruments. About 31.5\% of publications report XRT data without use of BAT or UVOT data. For 43.9\%, XRT data are used in combination with the other two instruments BAT and UVOT. The least used instrument is UVOT: 6.4\% alone, 28.1\% in combination with the others. BAT is used 17.3\% alone, 29.8\% with the others.

\begin{table*}
\caption[t3]{Use of instruments for \swift\ papers\label{tinstruments}}
\begin{center} 
\begin{tabular}{cccccccc}
\tableline\tableline&&&&&&&\\[-5pt] 
Year & BAT/XRT/UVOT & BAT/XRT & XRT/UVOT & BAT/UVOT & BAT & XRT & UVOT \\
[5pt]\tableline&&&&&&&\\[-5pt] 
2005 & 5 & 4 &  3 & 0 &  4 & 8 &  0 \\
2006 & 24 & 16 & 7 & 0 & 10 & 20 & 2 \\
2007	 & 18 & 34 & 18 & 1 & 19 & 57 & 5 \\
2008	 & 11 & 25 & 22 & 0 & 25 & 44 & 16 \\
2009	 & 20 & 32 & 33 & 3 & 35 & 56 & 11 \\
2010 & 30 & 37 & 33 & 2 & 46 & 64 & 12 \\
2011 & 31 & 34 & 47 & 2 & 51 & 98 & 24 \\
[2pt]\tableline
Fraction & 12.6\% & 16.5\% & 14.8\% & 0.7\% & 17.3\% & 31.5\% & 6.4\% \\
[2pt]\tableline
\end{tabular}
\end{center}
\end{table*}

\section{Conclusions}\label{discussion}

Our study confirms that, along with large observatories, it is extremely important for science to invest in and build medium- and small-size missions, and support them for the longest possible time. \swift\ is an international small high-energy mission, launched in 2005 and operated by a team of 15 scientists. Including launch and 8 years of operation, it cost 300 million USD.  The success of \swift\ has already been demonstrated by Madrid \& Macchetto (2009), who also point out that the most successful observational project of the last decade is the Sloan Sky Digital Survey and its 2m telescope. \swift\ showed early on that it would be a success, with a total of 24 papers in 2005. In 2011, it produced 287 papers, much more than Gemini, doubling Subaru, and reaching one of the most successful observatories ever, Keck. Citations per paper are around the average for the other observatories, with Keck generally achieving the highest citation rate. It is clear that saturation for \swift\ might very likely not yet have been reached in 2011 and we expect the number of publications for 2012 to be higher than in the year before. 

Similar conclusions are reached by Abt (2012), who indicates that the extra costs of large ground-based telescopes are not totally compensated by the higher number of citations. This comparison between large and small facilities relates to recent  conclusions by Loeb (2012): most important discoveries occur serendipitously, and serendipity must go together with large planned projects. Similarly, if we consider the means used to make discoveries, we argue that successful small missions are very important, should continue to be supported as long as possible, and should not be regarded as a competition for large ones. 

\acknowledgments
We thank Neil Gehrels,  John Nousek and Julian Osborne for valuable information about \swift, and Jill Lagerstrom and Patricia Schady for helpful comments. It is a pleasure to acknowledge Sarah Stevens-Rayburn for carefully reading the manuscript. This research has made use of NASA's Astrophysics Data System.


\begin{thebibliography}{}


\bibitem[Abt (2012)]{Abt} Abt, H.~A.\ 2012, AJ, 144, 91
 
\bibitem[Apai et al.(2010)]{2010PASP..122..808A} Apai, D., Lagerstrom, J., 
Reid, I.~N., et al.\ 2010, \pasp, 122, 808 

\bibitem[Barthelmy(2005)]{Barthelmy} Barthelmy S.~D., et al.\ 2005, Space Science Review, 120, 143

\bibitem[Burrows(2005)]{Burrows} Burrows D.~N., et al.\ 2005, Space Science Review, 120, 165

\bibitem[Conley \& Stadmark(2012)]{Conely} Conley D., \& Stadmark J.\ 2012, \nat, 488, 590

\bibitem[Erdmann \& Grothkopf(2010)]{erd10} Erdmann, C., \& Grothkopf, U.\ 2010, Library and Information Services in Astronomy VI: 21st Century Astronomy Librarianship, From New Ideas to Action, 433, 81

\bibitem[Evans et al.(2009)]{2009MNRAS.397.1177E} Evans, P.~A., Beardmore, A.~P., Page, K.~L., et al.\ 2009, \mnras, 397, 1177 

\bibitem[Gehrels et al.(2004)]{2004ApJ...611.1005G} Gehrels, N., 
Chincarini, G., Giommi, P., et al.\ 2004, \apj, 611, 1005

\bibitem[Grothkopf 
\& Meakins(2012)]{2012Msngr.147...41G} Grothkopf, U., \& Meakins, S.\ 2012, The Messenger, 147, 41 


\bibitem[Loeb(2012)]{2012arXiv1207.3812L} Loeb, A.\ 2012, arXiv:1207.3812 

\bibitem[Madrid 
\& Macchetto(2009)]{2009BAAS...41..913M} Madrid, J.~P., \& Macchetto, D.\ 2009, \baas, 41, 913 

\bibitem[Meakins \& Grothkopf(2012)]{meak11} Meakins, S., \& Grothkopf, U.\ 2012, Astronomical Data Analysis Software and Systems XXI, 461, 767 

\bibitem[Roming(2005)]{}  Roming P.~W.~A., et al.\ 2005, Space Science Review, 120, 95

\bibitem[Rots et al.(2012)]{2012PASP..124..391R} Rots, A.~H., Winkelman, 
S.~L., \& Becker, G.~E.\ 2012, \pasp, 124, 391

\end{thebibliography}
\end{document}